# Gigacount/second photon detection with InGaAs avalanche photodiodes

K. A. Patel, J. F. Dynes, A. W. Sharpe, Z. L. Yuan, R. V. Penty and A. J. Shields

We demonstrate high count rate single photon detection at telecom wavelengths using a thermoelectrically-cooled semiconductor diode. Our device consists of a single InGaAs avalanche photodiode driven by a 2 GHz gating frequency signal and coupled to a tuneable self-differencing circuit for enhanced detection sensitivity. We find the count rate is linear with the photon flux in the single photon detection regime over approximately four orders of magnitude, and saturates at 1 gigacount/s at high photon fluxes. This result highlights promising potential for APDs in high bit rate quantum information applications.

*Introduction*: Single photon detection has become increasingly important for a wide range of applications, in particular quantum communication and quantum information processing [1]. The ultimate performance of these applications relies on the ability to detect a photon for each system clock cycle, *i.e.,* requiring a count rate up to 1 gigacount/sec for a system that is clocked at GHz frequencies. To date, a gigacount/sec rate has been reported only for liquid He temperature superconducting nanowire detectors [2], although here this count rate may have been achieved at the expense of the detection efficiency [3].

Semiconductor avalanche photodiodes (APDs) have been favoured for single photon detection [1] at a wavelength spectrum covering both visible and near-infrared, because they are compact and rugged, operate at close to room temperature, use little energy, and can be mass produced. The maximum count rate of commercial devices is restricted to ~10 MHz, due to slow device recovery following a successful detection event. In the case of InGaAs APDs, this recovery is more problematic due to afterpulsing, which is an additional source for error besides dark counts. To suppress afterpulse noise, an InGaAs APD is usually gated electrically to be single photon sensitive only during the expected arrival time of the photons.

Operating detectors in gated mode has now become an avenue to increase the bit rate for single photon applications, thanks to the significant evolution of this technology over the past decade [4-9]. We have developed a self-differencing (SD) circuit [7], which compares the APD output with its identical copy but temporally translated by an integer number of clock cycles, removing the periodical capacitive response of the APD and thus allowing detection of extremely weak avalanches which would otherwise be obscured. InGaAs SD-APDs have been used for the first 1-Mbit/s quantum key distribution system [10], and their maturity for practical use has been rigorously examined in the field environment with remote synchronisation [11].

Although InGaAs APDs can be gated at multi-GHz frequencies [8], it remains unclear whether they can also count single photons close to their analogue bandwidth, *i.e.,* with a rate of 1 GHz or above. In this letter, we demonstrate for the first time gigacount per second rates using an InGaAs APD. The advance stems mainly from the enhanced sensitivity after incorporating tuneability [8].

*Experimental setup*: The InGaAs APD under study is thermo-electrically cooled to -30 ºC. The APD is gated by a 2 GHz, 5.2 V square wave, in combination with a DC bias of 52.3 V, to periodically stimulate the detector into a single photon sensitive state. To eliminate the capacitive response of the APD to this AC bias, which will obscure weaker avalanches, a tuneable SD circuit [8] is used to process the APD signal output, which is then followed by amplification and discrimination. In the detector characterisation, a 50-ps laser pulse at a wavelength of 1550 nm is incident every 64$^{th}$ gate period, *i.e.* at a repetition frequency of 31.25 MHz, on the APD. The photon flux is set to 0.2 photons/pulse using a calibrated variable optical attenuator and power meter. At 2 GHz, the detection efficiency is characterized to be 20% with an afterpulse probability of 4% and dark count probability of 1.8×10$^{-5}$ per gate. The measured timing jitter of the detector is 170 ps.

*Self-differencing performance:* Figure 1 shows the performance of the SD circuit in the frequency domain. The spectra show the raw (solid line) and the SD (dashed line) outputs respectively. The raw output is strongly dominated by the capacitive response which masks the weaker avalanche signals, as evidenced by the strong power peak centred at 2 GHz. After the SD circuit, the signal power at 2 GHz is suppressed strongly by approximately 61 dB, leaving a small residual background as shown in the inset of Fig. 2. This suppression allows the sensing of avalanches of charge as low as 0.035 pC [8]. This sensitivity is significant improvement over the previous SD circuit without tuneability, which senses an average charge of 0.2 pC per avalanche [12].

*Results and discussion:* To measure the maximum count rate, the detector is illuminated with a 15-ps laser pulsed at 1 GHz, a frequency that is exactly half of the gating frequency to the APD. This illumination arrangement avoids unwanted cancellation between photon-induced avalanches. The APD output is captured by an oscilloscope of 4 GHz analogue bandwidth, and the photon count rate determined from the number of times the waveform rises above the discrimination level. The count rate determined by the oscilloscope agrees closely with that measured directly with photon counting electronics, up to a count rate of 4 megacounts/sec limited by the linear regime of the latter technique.

Figure 2 shows the count rate as a function of incident photon flux. Throughout the measurement, the discrimination level is fixed at the level shown by the dotted line in the inset. Three distinctive regions have been identified: (i) at low fluxes the count rate is dominated by the dark counts; (ii) in the single photon regime, the count rate is linear over approximately 4 orders of magnitude of the incident flux; and (iii) at higher photon fluxes, the count rate saturates at 1 GHz. The 1 gigacount/s rate is more visible in the inset of Fig. 2: there is a clean avalanche signal in every 1 ns window in the waveform recorded by the oscilloscope.

The measured count rate ($R$) can be simulated by the equation

$$R = f_0[(1-e^{\mu\eta}) + P_d - (1-e^{\mu\eta})P_d],$$

where $f_0$ is the illumination clock frequency, $\mu$ the photon flux, $\eta$ the single photon detection efficiency and $P_d$ the dark count probability. In Fig. 2, excellent agreement can be found between the measured data and theoretical simulation. A constant $\eta$ of 20% is used in the simulation, suggesting that the single photon detection efficiency is independent of the count rate. This constant efficiency is attributed to the low avalanche current. At the saturated 1 GHz count rate, the measured current is 233 µA, which is sufficiently low so as not to heat the device or change the bias condition significantly.

We remark that the demonstrated 1 GHz count rate is the highest among any semiconductor based single photon detectors, including multi-pixel devices. Importantly, this count rate is achieved without any sacrifice in the single photon detection efficiency. This achievement is attributed to two factors. First, the enhanced sensitivity of the SD circuit and hence lower avalanche charge make it possible for rapid recovery of the detector after an avalanche event. Second, the SD circuit generates an avalanche signal which is composed of both positive and negative components in equal measure (see Fig. 2, inset). This provides a balanced output which does not give rise to any DC signal drift, thus allowing the discrimination level to be set independent of count rate. This feature also explains the count-rate independent timing jitter [12], which is of paramount importance for high bit rate applications.

*Conclusion:* The count rate of InGaAs APDs has now been improved to 1 gigacount/s, comparable to their analogue bandwidth. This result highlights promising potential of semiconductor APDs to further increase the bit rates in single photon applications such as quantum key distribution [9] and quantum random number generation [13].

———

*Acknowledgement:* Partial support from the UK Technology Strategic Board under contract 400207 and the EU FP7 Integrated Project Q-Essence is acknowledged. K. A. P acknowledges personal support via the EPSRC funded CDT in Photonic System Development.

**Authors' affiliations:**

K. A. Patel (*Toshiba Research Europe Ltd., Cambridge Research Laboratory, 208 Cambridge Science Park, Cambridge, CB4 0GZ, United Kingdom and Cambridge University Engineering Department, 9 JJ Thomson Ave, Cambridge, CB3 0FA, United Kingdom*)

J. F. Dynes, A. W. Sharpe, Z. L. Yuan and A. J. Shields (*Toshiba Research Europe Ltd., Cambridge Research Laboratory, 208 Cambridge Science Park, Cambridge, CB4 0GZ, United Kingdom*)

R. V. Penty (*Cambridge University Engineering Department, 9 JJ Thomson Ave, Cambridge, CB3 0FA, United Kingdom*)

**Figures:**

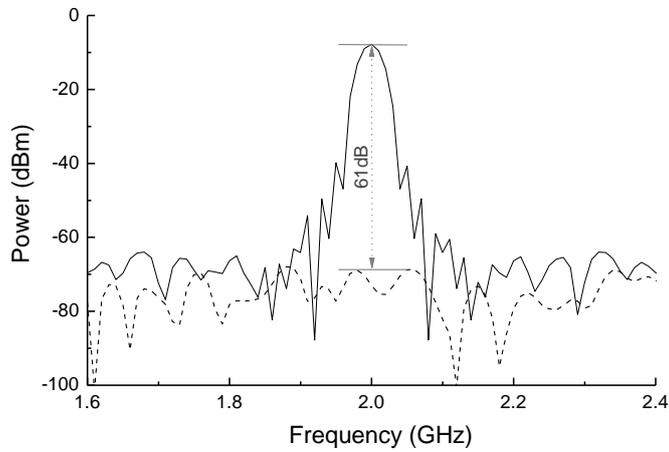

Fig. 1 Frequency power spectrum for the raw APD output and the SD output.

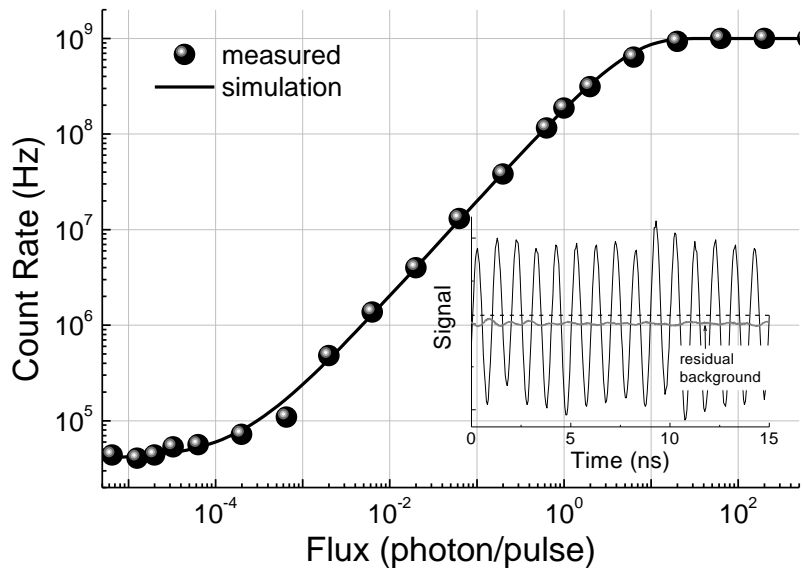

Fig. 2 Photon count rate as a function of the incident photon flux. The solid line shows the theoretical simulation using Eq. (1) with a single photon detection efficiency of 20%. Inset: a waveform recorded for the APD under an illumination flux of 50 photons per pulse (solid black line) as well the residual background recorded under no illumination (thick grey line), respectively. The dashed line indicates the discrimination level set for all incident photon fluxes.